\documentstyle[12pt]{article}

\def\ai{\'{\i}}
\begin{document}
\baselineskip.33in

\centerline{\large{\bf Global phase time and wave function}}

\smallskip

\centerline{\large{\bf for the Kantowski--Sachs anisotropic universe}}

\bigskip

\centerline{  Claudio Simeone\footnote{{\bf Electronic mail:} simeone@tandar.cnea.gov.ar}}

\medskip

\centerline{\it Departamento de F\ai sica, Comisi\'on Nacional de Energ\ai a At\'omica}

\centerline{\it Av. del Libertador 8250, 1429 Buenos Aires, Argentina}

\centerline{\it and}

\centerline{\it Departamento de F\ai sica, Facultad de Ciencias Exactas y Naturales}

\centerline{\it Universidad de Buenos Aires,  Ciudad Universitaria, Pabell\'on I}
\centerline{\it 1428, Buenos Aires, Argentina.}

\vskip1cm

\noindent ABSTRACT

\bigskip
 
\noindent  A consistent quantization  with a clear notion of time and evolution is given for the  anisotropic Kantowski--Sachs cosmological model.  It is shown that a suitable coordinate choice allows to obtain  a solution of the Wheeler--DeWitt equation  in the form of definite energy states, and that  the results can be associated to two disjoint equivalent theories, one for each sheet of the constraint surface.

\vskip1cm


\vskip1cm


\newpage

\noindent{\bf 1. Introduction}

\bigskip

\noindent The difficulty in defining a set of observables and a notion of dynamical evolution in a theory where the spacetime metric is itself a dynamical variable, as it is the case of General Relativity, leads to the problem of time in quantum cosmology. The time parameter $\tau$ entering the formalism is not a true time, and as a consequence of this, the theory  includes a constraint ${\cal H}\approx 0$ [1,2].
In the Dirac--Wheeler--DeWitt canonical quantization of minisuperspace models one introduces a wave function $\Psi$ which must fulfill the operator form of the constraint equation, that is,
\begin{equation}
{\cal H}\Psi=0,
\end{equation}
where the momenta are replaced in the usual way by  operators  in terms of derivatives of the coordinates:
$$p_k=-i{\partial\over\partial q^k}.$$
 As the Hamiltonian is quadratic in $p_k$ a second order differential equation is obtained; this is called the Wheeler--DeWitt equation [3].   It is clear that the solution $\Psi$ does not depend explicitly on the time parameter $\tau$, but only on the coordinates $q^k$. This is the main problem with the Dirac--Wheeler--DeWitt quantization, because the absence of a clear notion of time makes  difficult to have a definition of conserved positive-definite probability, and therefore to guarantee the unitarity of the theory. To build the space of physical states we need to define an inner product which takes into account that there can be a physical time ``hidden'' among the canonical variables of the system. The physical inner product $(\Psi_2|\Psi_1)$ must be defined by  fixing the time in the integration.
If the time can be defined  as $t(q)$ ({\it intrinsic time}) then we can introduce an  operator  ${\hat\mu}_{t'} =\delta (t(q)-t')$, which evaluates the product at the fixed time $t'$. Hence, to obtain a closed theory by this way we need a formally correct definition of time.
In fact, if we have been able to isolate the time as a function of the canonical variables, we could work with the new coordinates $(t,q^\gamma)$ and the corresponding momenta $(p_t,p_\gamma)$ and make the  substitution $p_t=-i\partial/\partial t$,  
$p_\gamma=-i\partial/\partial q^\gamma$ to obtain a Wheeler--DeWitt equation whose solution will depend on $t$, so that it will have an evolutionary form. 

In a previous work we gave a proposal for quantizing the Kantowski--Sachs anisotropic universe with a clear notion of time within the path integral formulation [4]. The procedure was based on a canonical transformation which turned the action of the minisuperspace into that of an ordinary gauge system, which allowed to use canonical gauge conditions to identify a  global phase time [5,6]. A not completely satisfactory point was the form of the resulting propagator: the  expression obtained could not be explicitly integrated, so that the interest of the method was almost purely formal. In the present work, instead,  we obtain  a  consistent quantization with a right notion of time and evolution within the canonical formalism; we give an explicit form for the wave function by solving a Wheeler--DeWitt equation in terms of coordinates including a global time and such that the reduced Hamiltonian is time-independent, so that the result can be understood as a set of definite energy states. In the Appendix it is shown that the time here employed in the Wheeler--DeWitt quantization can be obtained with our deparametrization procedure proposed in Ref. 4.

\vskip1cm

\noindent{\bf 2. The Kantowski--Sachs universe}

\bigskip

\noindent Possible anisotropic cosmologies are comprised by the Bianchi models and the Kantowski--Sachs model [7,8,9].
 By introducing the diagonal $3\times 3$ matrix $\beta_{ij}$, anisotropic  spacetime metrics can be put in the form
\begin{equation}
ds^2= N^2d\tau^2-e^{2\Omega(\tau)}(e^{2\beta(\tau)})_{ij}\sigma^i\sigma^j,
\end{equation}
with the differential forms $\sigma^i$  fulfilling
$d\sigma^i=\epsilon_{ijk}\sigma^j\times\sigma^k$ [10]. The Kantowski--Sachs model corresponds to the case
$$\beta_{ij}=diag(-\beta,-\beta, 2\beta)$$ 
 $$\sigma^1=d\theta,\ \ \ \  \ \  \sigma^2=\sin\theta d\varphi, \ \ \ \ \ \ \ \sigma^3=d\psi$$
so that
\begin{equation}ds^2=N^2d\tau^2-e^{2\Omega(\tau)}\left(e^{2\beta(\tau)} d\psi^2+e^{-\beta(\tau)}(d\theta^2+\sin^2\theta d\varphi^2)\right).\end{equation}
Differing from the Bianchi universes, the model is anisotropic even in the case  $\beta_{ij}=0$. If matter is neglected in the dynamics, the classical behaviour of this universe is analogous to that of the closed Friedmann--Robertson--Walker cosmology in the fact that the volume, defined as 
$${\cal V}=\int d^3 x \sqrt{-(^3g)},$$
grows to a maximun and then returns to zero.  This feature is reflected in the form of the scaled Hamiltonian constraint $H=e^{3\Omega}{\cal H}$:
 \begin{equation}
H= -\pi_\Omega^2+\pi_\beta^2-e^{4\Omega+2\beta}\approx 0,
\end{equation}
which clearly allows for $\pi_\Omega=0$. As a consequence of this,  no function of only the coordinate $\Omega$ can be a global phase time for the Kantowski--Sachs universe: the required condition $[t,H]>0$  [11] cannot be globally fulfilled by $t=t(\Omega)$. In some early works, $\Omega$ was defined as the time coordinate, but this yields a reduced Hamiltonian which is not real for all possible values of $\beta$ and $\pi_\beta$; hence the corresponding Hamiltonian operator would not be self-adjoint, and the resulting quantum theory would not be unitary.

\vskip1cm

\noindent{\bf 3. Global time and Wheeler--DeWitt equation}

\bigskip

\noindent The coordinate $\beta$ has a non vanishing Poisson bracket with the constraint (5), then allowing for its use as time variable (this is not true if a matter field is included, as the corresponding additional term $\pi_\phi^2$ in the constraint makes possible $\pi_\beta=0$); but this would lead to a reduced  Hamiltonian with a time-dependent potential, making a ``one particle'' interpretation of the wave function impossible [12]. Then we  change to a set of new variables defined as
\begin{eqnarray}
x& \equiv & 2\Omega+\beta\nonumber\\
y& \equiv & \Omega+2\beta,
\end{eqnarray}
and rescale  the Hamiltonian in the
following form 
$$
H\rightarrow  {1\over 3}H,
$$
and thus we obtain the equivalent constraint
\begin{equation}
H=-\pi_x^2+\pi_y^2-{1\over 3} e^{2x}\approx 0\label{h}.
\end{equation}
 The momentum   $\pi_y$ does not vanish on the constraint surface; then we have $[y,H]=2\pi_y\neq 0$ and, up to a sign determined by the sign of $\pi_y$, the coordinate $y$ is a global phase time:
\begin{equation}
t=y\, sign(\pi_y) .
\end{equation}
On the other hand, we have a reduced Hamiltonian on each sheet of the constraint surface: $h(x,\pi_x)=\pm\sqrt{\pi_x^2+{1\over 3} e^{2x}}$, and the potential does not depend on time. 

Now, let us define $u=\sqrt{ 1\over 3}e^x$. In terms of the coordinates $u$ and $y$ the Wheeler--DeWitt equation associated to the constraint (\ref{h}) has the  form
\begin{equation}
\left( u^2\frac{d^2}{du^2}+u\frac d{du}-u^2-\frac{d^2}{dy^2}%
\right) \Psi (u,y)=0  \label{wdw}
\end{equation}
This equation  clearly admits a set of solutions of the form $\Psi=A(u)B(y)$; the solution for $B(y)$ is inmediately obtained in the form of exponentials of  imaginary argument, while for $A(u)$ we have a modified Bessel equation. Returning to the variable $x$, the solutions can then be written as  
\begin{eqnarray}
\Psi _\omega (x,y) & = & \left[ a^{+}(\omega )e^{i\omega y}+a^{-}(\omega )e^{-i\omega y}\right]\nonumber\\
& & \times\left[ b^{+}(\omega )I_{i\omega }\left(\sqrt{1\over 3 }e^x\right)+b^{-}(\omega )K_{i\omega
}\left(\sqrt{1\over 3}e^x\right)\right],   \label{bessel}
\end{eqnarray}
 where $I_{i\omega}$ and $K_{i\omega}$ are modified Bessel functions.
The contribution of the functions $I_{i\omega}\left(\sqrt{1\over 3 }e^x\right)$ must be discarded, because they diverge in the classicaly forbidden region associated to the exponential potential ${1\over 3}e^{2x}$. Because the coordinate $y$ is a global time, we could  separate the functions in (\ref{bessel}) as positive and negative-energy solutions, each subset corresponding to one sheet of the constraint surface, that is, to one of both signs of the reduced Hamiltonian. As $\pi_y={2\over 3}\pi_\beta-{1\over 3}\pi_\Omega$ and on the constraint surface we have $|\pi_\beta|>|\pi_\Omega|$, then $sign(\pi_y)=sign(\pi_\beta)$ and each set of solutions corresponds to one of  both signs of the original momentum $\pi_\beta$. Going back to the original coordinates we can write the wave function(s) as
\begin{eqnarray}
\Psi^+_\omega(\Omega,\beta) & = & c^+(\omega )e^{i\omega (\Omega+2\beta)}K_{i\omega
}\left(\sqrt{1\over 3}e^{2\Omega+\beta}\right)\nonumber\\
\Psi^-_\omega(\Omega,\beta) & = & c^-(\omega )e^{-i\omega (\Omega+2\beta)}K_{i\omega
}\left(\sqrt{1\over 3}e^{2\Omega+\beta}\right).
\end{eqnarray}
Note, however, that Because $t=(\Omega+2\beta)\, sign(\pi_\beta)$,  we can give a single expression \begin{equation}\Psi_\omega(x=2\Omega+\beta,t)  =  c(\omega )e^{i\omega t}K_{i\omega
}\left(\sqrt{1\over 3}e^{x}\right)\end{equation}
 for both sheets of the constraint surface,  reflecting that both disjoint quantum theories
are equivalent.

\vskip1cm

\noindent{\bf 4. Comments}

\bigskip

\noindent We have given a consistent quantization with a right notion of time for the anisotropic Kantowski--Sachs universe; we have improved previous analysis [4], by giving and explicit form of the wave function, which results of a Wheeler--DeWitt equation with a coordinate as a global time. 
This allows for a right definition of the space of physical states; if the solutions  are given in terms of the time $t$, the Klein--Gordon  inner product  defined as [11]
$$(\Psi_1|\Psi_2)={i\over 2}\int_{t=const}dx\left[\Psi^*_1{\partial \Psi_2\over\partial t}-\Psi_2{\partial \Psi^*_1\over\partial t}\right],$$
 is conserved and  positive-definite. 
A point to be signaled is that we can also define the  inner product in the space of physical states as a Schr\"{o}dinger inner product, 
$$(\Psi_1|\Psi_2)=\langle\Psi_1|\hat\mu|\Psi_2\rangle$$
with $\hat\mu=\delta(y-y')$, 
 because the  
Hamiltonian $H$ admits a factorization in the form of a product of two  constraints linear in $\pi_y=\pm\pi_t$, each one leading to a 
Schr\"{o}dinger equation [13,14]. This is possible because in terms of the new coordinates $(x,y)$ the potential  does  not depend on time, and therefore $\pi_t$ conmutes with the reduced  Hamiltonian $h$. Recall that such a factorization is not possible if we work with the original coordinates $\Omega,\beta$; in this case $t=\pm\beta$, and $\pi_t$ does not conmute with the (time-dependent) reduced Hamiltonian $h=\pm\sqrt{\pi_\Omega^2+e^{4\Omega+2\beta}}$. 

It should be noted that we have started from a scaled constraint $H=e^{3\Omega}{\cal H}$, which  at the classical level is equivalent to ${\cal H}$, but which does not necessarily lead to the same quantum description. However, it can be shown that an operator ordering exists such that both constraints $H$ and ${\cal H}$ are equivalent at the quantum level. Let us consider a generic constraint
$$
e^{bq_1}\left(-p_1^2+p_2^2+\zeta e^{aq_1+cq_2}\right)\approx 0,
$$
which contains an ambiguity associated to the fact that the most general form of the first term should  be written 
$$-e^{A q_1} p_1e^{(b-A-C) q_1} p_1e^{C q_1}$$
(so that $A$ and $C$ parametrize all possible operator orderings). It is simple to verify that the constraint with the most general ordering differs from that with the trivial ordering in two terms, one linear and one quadratic in $\hbar$, and that these terms vanish with the choice $C=b-A=0$. Therefore, the Wheeler--DeWitt equation resulting from the scaled constraint $H$ is right in the sense that it corresponds to a possible ordering of the original constraint ${\cal H}$.

A final remark should be that the obtention of two disjoint theories corresponding to both sheets of the constraint surface is possible because the model under consideration admits an intrinsic time: the time is given by the coordinate conjugated to the non vanishing momentum whose sign identifies each sheet. In the case of a model such that all the momenta could vanish, the procedure could still be carried out if a canonical transformation  leading to a non vanishing potential can be performed [15]. 

\vskip1cm

\noindent{\bf Acknowledgement}

\bigskip

\noindent I wish to thank G. Giribet and F. D. Mazzitelli for very helpful comments.

\vskip1cm

\noindent{\bf Appendix}

\bigskip

\noindent In Ref. 4 we proposed a deparametrization procedure based in a canonical transformation turning the minisuperspace into an ordinary gauge system, so that a $\tau$-dependent canonical gauge condition could be used to define a global phase time [5,6]. The generating function of the transformation must be a solution of the Hamilton--Jacobi equation
$$H\left(q^k,{\partial W\over\partial q^k}\right)=E$$
which results of matching the $E$ to one of the new momenta, for example $\overline P_0$. Therefore the Poisson bracket of the new coordinate $\overline Q^0$  with the constraint $H$ is equal to unity, and $\overline Q^0$ can be used to fix the gauge. In Ref. 4 we started from a different expression for the constraint, and the resulting (intrinsic) time had the form of an exponencial of $\Omega$ and $\beta$. 

It is simple to verify that this method can reproduce the definition of time given  here. If we start from the constraint (6) we obtain the Hamilton--Jacobi equation 
$$-\left(\partial W\over\partial x\right)^2+ \left(\partial W\over\partial y\right)^2-{1\over 3}e^{2x}=E.$$
Introducing the integration constants $\overline P_0=E$ and $a$ such that $a^2+E=\pi_y^2$ we obtain the solution
$$W=y\, sign(\pi_y)\sqrt{a^2+\overline P_0}+sign(\pi_x)\int dx\sqrt{a^2-{1\over 3}e^{2x}}$$
so that
$$\overline Q^0={\partial W\over\partial\overline P_0}={y\,sign(\pi_y)\over 2 \sqrt{a^2+\overline P_0}}.$$
Then we can fix the gauge by means of the canonical condition $\chi\equiv 2\overline Q^0 \sqrt{a^2+\overline P_0}-T(\tau)=0$ with $T$ a monotonous function of $\tau$, which yields  $t=y\, sign(\pi_y)$. We could also define a time including the momenta ({\it extrinsic time}) by fixing $\chi\equiv 2\overline Q^0 -T(\tau)=0$, which gives $t=y/\pi_y$.

\newpage
\noindent{\bf References}

\vskip1cm

\noindent 1.  Barvinsky  A. O., Phys. Rep. {\bf 230}, 237 (1993).

\noindent 2. Ferraro R., Grav. Cosm. {\bf 5},  195 (1999).

\noindent 3. Halliwell  J. J., in {\it Introductory Lectures on Quantum Cosmology}, Proceedings of the Jerusalem Winter School on Quantum Cosmology and Baby Universes, edited by T. Piran, World Scientific, Singapore (1990). 

\noindent 4. Simeone  C., Gen. Rel. Grav. {\bf 32}, 1835 (2000).

\noindent 5. De Cicco H. and   Simeone C., Gen. Rel. Grav. {\bf 31}, 1225 (1999).

\noindent 6. Simeone  C., J. Math. Phys. {\bf 40}, 4527 (1999).

\noindent 7. Ryan M. P. and  Shepley L. C., {\it Homogeneous Relativistic Cosmologies},
Princeton Series in Physics, Princeton University Press, New Jersey (1975).

\noindent 8. Kantowski R. and Sachs R. K., J. Math. Phys.  {\bf 7}, 443 (1966).

\noindent 9. Higuchi  A. and  Wald R. M., Phys. Rev.  {\bf D51}, 544 (1995)

\noindent 10. Schutz B.  F., {\it Geometrical Methods of Mathematical Physics,}
Cambridge University Press, Cambridge (1980).

\noindent 11. H\'aj\'{\i}cek P., Phys. Rev.  {\bf D34}, 1040 (1986).

\noindent 12.  Kucha\v r  K. V., in {\it Quantum Gravity 2: A Second Oxford Symposyum}, edited by C. J. Isham, R. Penrose and D. W. Sciama, Clarendon Press (1981).

\noindent 13. Cavagli\`a M.,  De Alfaro V. and  Filippov A. T., Int. J. Mod. Phys. {\bf A10}, 611 (1995).

\noindent 14.   Catren G. and  Ferraro R., Phys. Rev. {\bf D63}, 023502 (2001).

\noindent 15. G. Giribet and C. Simeone,  Int. J. Mod. Phys. {\bf A17}, 2885 (2002)).

\end{document}